\documentclass[%
 aip,
 jap,%
 amsmath,amssymb,
 reprint,%
]{revtex4-1}

\draft

\usepackage{graphicx}
\usepackage{dcolumn}
\usepackage{bm}
\usepackage{url}
\usepackage{epstopdf}
\usepackage{epsfig}
\usepackage{color}
\usepackage[colorlinks]{hyperref}
	\hypersetup
	{
		colorlinks,%
		citecolor=black,%
		linkcolor=blue,%
		urlcolor=black,%
		filecolor=blue
	}
\DeclareGraphicsExtensions{.eps}

\begin{document}

\preprint{AIP/123-QED}

\title{First Principles Study of Metal Contacts to Monolayer Black Phosphorous}

\author{Anuja Chanana and Santanu Mahapatra}
\noaffiliation
\affiliation{ Nano-Scale Device Research Laboratory, Department of Electronic Systems Engineering, Indian Institute of Science (IISc) Bangalore, Bangalore 560012, India
}%
\date{\today}

\begin{abstract}
Atomically thin layered black phosphorous (BP) has recently appeared as an alternative to the transitional
metal di chalcogenides for future channel material in a MOS transistor due to its lower carrier effective mass. Investigation of the electronic property of source/drain contact involving metal and two-dimensional material is essential as it impacts the transistor performance. In this paper we perform a systematic and rigorous study to evaluate the Ohmic nature of the side-contact formed by the monolayer BP (mBP) and metals (gold, titanium and palladium), which are commonly used in experiments. Employing the Density Functional Theory (DFT), we analyse the potential barrier, charge transfer and atomic orbital overlap at the metal-mBP interface in an optimized structure to understand how efficiently carriers could be injected from metal contact to the mBP channel. Our analysis shows that gold forms a Schottky contact with a higher tunnel barrier at the interface in comparison to the titanium and palladium. mBP contact with palladium is found to be purely Ohmic, where as titanium contact demonstrates an intermediate behaviour.
\end{abstract}

\maketitle

\section{\label{sec:level1}Introduction}

Two dimensional materials have appeared as replacements for bulk Silicon in future MOS (metal-oxide-semiconductor) transistor  channel, \cite{doi:10.1021/nl025639a, tans1998room, Martel1998, duan2003high, doi:10.1021/nl025875l, Lu2008, Obradovic2006a, doi:10.1021/nl803176x} as they can offer the ultimate electrostatic integrity to scale down technology roadmap in sub-decananometer regime. Though the carrier mobility is extremely high in pristine graphene, \cite{bolotin2008ultrahigh} due to its the near-zero bandgap, its practical application as an electronic switch is questionable. \cite{Schwierz2010, Sordan2009} Since the pioneering work by Radisavljevic et al.\cite{Radisavljevic2011}, transitional metal dichalcogenides (TMDs), which inherit intrinsic band-gap comparable to Silicon, have attracted much attention \cite{doi:10.1021/nn501013c, Levi2013, Das2013a, doi:10.1021/nl301702r, larentis2012field, ADMA:ADMA201305845, liu2011performance} for MOS transistor channel application. Apart from the electrostatic integrity (which controls the OFF state current), the ON current provided by a transistor is another important parameter for technology scaling. Due to high value of the effective mass, the ON current of TMD-based MOS transistors are found to be low. Very recently, significantly higher ON current based MOS transistor is demonstrated experimentally, which uses atomically thin few layers black phosphorous (BP)\cite{Li2014} instead of TMDs. Monolayer-BP coined as phosphorene \cite{Li2014}, has an orthorhombic structure and inherits lower effective mass than the MoS$_2$. It is the least reactive hexagonal allotrope of elementary phosphorous with puckered layers of atoms stacked to each other by weak vander Waals (vdW) interactions.\cite{du2010ab} Each atom is covalently bonded to three other atoms. First principles based calculation demonstrates the direct band gap  in the range of 0.91 eV - 0.28 eV for mono to five layer BP.\cite{Qiao2014} Thickness dependent and highly anisotropic field effect mobility values upto 1000 cm$^{2}$ V$^{-1}$ s$^{-1}$ are achieved using few layer BP FET\cite{Li2014}. Contrary to the multilayer TMDs, which have indirect band gap\cite{Ellis2011}, multilayer BP shows a direct band gap that is advantageous to electronic and optoelectronic applications\cite{PhysRevB.89.235319,rodin2014strain,low2014tunable}.

Deficiency of effective doping techniques poses a serious challenge in attaining purely Ohmic contacts in such transistors, as it limits the current flow in the two dimensional (2D) channel. Potential barrier to charge carries at the metal-semiconductor interface needs to be reduced efficiently to achieve minimum contact resistance. Hence, a proper choice of contact metals which can form Ohmic contacts to 2D materials is an essential metric. Therefore, a careful examination is required which at the same time can also maximize carrier injection efficiency. 

Previous studies consider Cu(111), Zn(0001), In(110), Ta(110) and Nb(110) as suitable contact materials for mBP on the basis of minimum lattice mismatch and conclude Cu(111) as the best choice among all.\cite{PhysRevB.90.125441} Ferromagnetic tunnel contacts TiO$_{2}$/Co with tunable resistance is reported too. \cite{kamalakar2014nanolayer} In this work we choose Gold (Au)\cite{Li2014}, Titanium (Ti)\cite{Li2014} and Palladium (Pd),\cite{Xia2014,doi:10.1021/nn502553m} which are commonly used as contact materials, to assess the Ohmic nature of the metal-mBP interface through first principles Density Functional Theory (DFT). We first build a metal-mBP interface and obtain an optimized inter layer separation using conventional binding energy calculations. Then we study the minimum barrier potential, maximum electron density and maximum orbital overlap in optimized interface geometries to find the best Ohmic nature in gold, titanium and palladium contacts using DFT. Eventually it is observed that Pd offers an Ohmic contact (best carrier injection) and Au provides a Schottky contact (worst carrier injection), while Ti shows a nature close to Ohmic contact.

\section{\label{sec:level1}Computational Details and Methods}
First Principles DFT calculations based on conventional Kohn Sham\cite{PhysRev.136.B864, PhysRev.140.A1133} Hamiltonian are performed using Perdew Burke Ernzerhof Generalized Gradient Approximation GGA-PBE\cite{PhysRevB.54.16533, perdew1996generalized} as implemented in Atomistix Tool Kit (ATK).\cite{QumWS} Previously reported optimized lattice constants for mBP (a = 3.32\AA{} b = 4.58\AA{})\cite{Qiao2014} and experimental lattice constants for Au,\cite{PSSB:PSSB19630030312} Ti\cite{wood1962lattice} and Pd\cite{dutta1963lattice} are considered for the present study. Fritz Haber Institute (FHI) pseudopotential using Troullier-Martins\cite{PhysRevB.43.1993} scheme having a double zeta polarized basis sets are employed for phosphorous and each contact metal. Pulay mixer algorithm is used as the iteration control parameter with the maximum number of iteration steps as 100 and a tolerance value upto 10$^{-5}$ Hartree. Fast Fourier transform is utilized as the Poisson Solver. Monkhorst Pack\cite{monkhorst1976special} k-point sampling is set as 8x8x1 along with a density mesh cut off of 90 Hartree. The parameters are kept same for simulations of individual material (semiconductor and the contact metal) and for the interface formed by the two. As aforementioned optimized lattice parameters are considered for mBP while the contact structure are relaxed until the forces on every atom is 0.01 eV/\AA{}.  A supercell is formed containing 6 layers (reported earlier)\cite{giovannetti2008doping} of metal surface interfaced with 1x1 unit cell of mBP. A vacuum region of more than 30\AA{} is keep between periodic structure of metal-mBP interface to avoid spurious interactions between  the individual structures.

For the metal contact evaluation the analysis is divided in two steps as outlined in Fig 1. The first step aims in building an optimized geometry of metal-mBP.  For this optimized lattice parameters of each metal and mBP are considered and then the interfacial strain is calculated. The estimation of the optimized interface geometry is done using the binding energy(BE) calculation and the distance with minimum BE is the optimized inter layer distance. The second step involves systematic calculation of potential barrier using effective potential, electrostatic difference potential, electron localization function; charge transfer using electron density, electron difference density and mulliken population; and orbital overlap using density of states and band structure all at the interface using DFT simulations.  All the above analysis collectively is used to assess the effective carrier injection from contact to channel.   
\begin{figure}
\includegraphics[width =1.0\columnwidth] {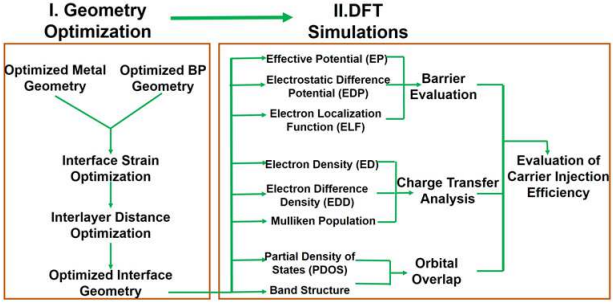}
\caption{\label{fig1}Algorithm used for the metal-mBP contact evaluation.} 
\end{figure}

\section{\label{sec:level1}Results and Discussions}
\subsection{\label{sec:level2}Band structure Analysis of Monolayer Black Phosphorous}
BP, the most stable form of phosphorus at room temperature was first discovered by Bridgman\cite{doi:10.1021/ja02184a002} way back in 1914 from ordinary white phosphorous. It has individual layers with puckered honeycomb structure as shown in Fig 2(a). The side and top views of mBP is presented in Fig 2(b), (c) and (d). The crystal structure of BP is orthorhombic with Cmca No.64 Space Group. The primitive cell is made of four atoms with intra layer bindings as covalent and inter layer interactions as vander Waals. Fig 2 (e) and (f) shows the bandstructure (high symmetry points G,X,S,G,Y,S) and Broullin Zone of mBP. The lattice parameters (a = 3.32\AA{} b = 4.58\AA{})\cite{Qiao2014} reported previously are considered for present study. The band gap values using GGA-PBE\cite{perdew1996generalized} and MGGA-TB09\cite{perdew1999accurate, Tran2009} obtained are 0.93 eV and 1.41 eV at the gamma point of BZ which are in close agreement with preceding articles.\cite{Qiao2014, doi:10.1021/nn501226z}. The MGGA results is found to be in near equivalence to HSE reported band gap values. This verification is found to be  consistent with the previously reported values of band gaps of various materials calculated using MGGA.\cite{PhysRevB.82.205212}
\begin{figure}
\includegraphics[width =1.0\columnwidth] {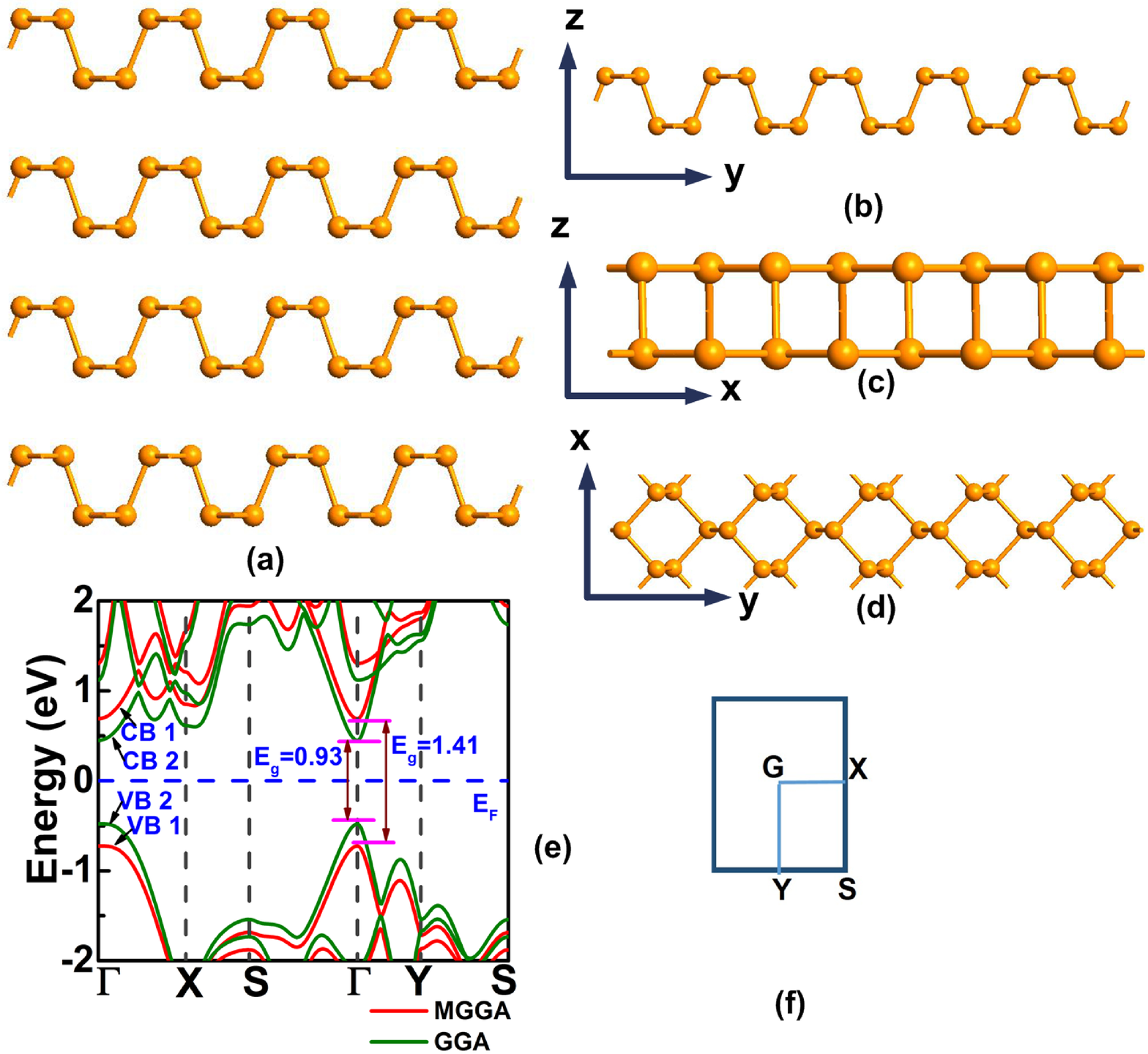}
\caption{\label{fig1}(a) Puckered honeycomb layers of Black Phosphorous (b),(c) and (d) Side and Top view of mBP (e) Comparison of band structure obtained using GGA and MGGA, the band structure lines are same for both MGGA and GGA, just the MGGA lines are shifted upwards showing a higher value of band gap (E$_{g}$) as compared to GGA at the gamma point, E$_{F}$ is the Fermi Level. (f) Brillouin Zone of mBP.}
\end{figure}
\subsection{\label{sec:level2}Contact Geometry and Lattice Matching}
After the band gap analysis of mBP, we proceed further with metal-mBP contact evaluation. The algorithm we followed is outlined in Fig 1. As per the first step detailed in the algorithm, we consider the optimized geometry of each of metal contacts (Au, Ti and Pd) as well as the optimized lattice parameters for mBP.  The work function of  semiconducting mBP is calculated to be 4.6 eV which is in close agreement to the previously reported value. \cite{PhysRevB.90.125441,cai2014layer}The work function of mBP and work function each of the metals\cite{Michaelson1977} along with the top and side view of side contact geometry\cite{doi:10.1021/nn405249n}  with mBP as the channel material is shown in Fig 3 (a), (b) and (c). The optimized structures for all the metal contacts are obtained first before interfacing them with mBP.
\begin{figure}
\includegraphics[width =1.0\columnwidth] {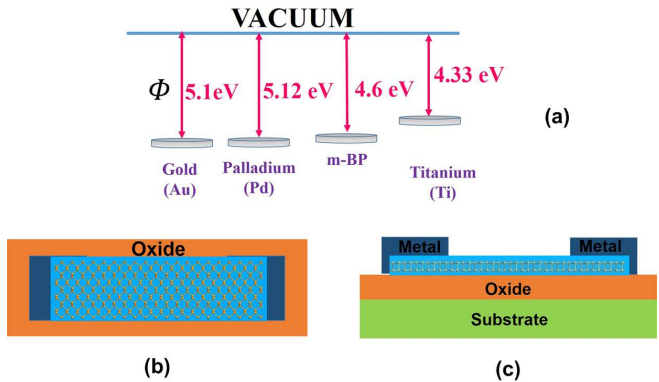}
\caption{\label{fig1}(a) Energy diagrams showing Work Function of mBP and all metals. Schematic illustration of (b) Top View and (c) Side View of Side Contact geometry with mBP as the channel material  }
\end{figure} 

(111) cleaved surface for FCC (Au and Pd) and (0001) for HCP (Ti) lattice structure are interfaced with mBP. The lattice constants of mBP are taken from earlier reports, hence it is found appropriate to strain the metal layers w.r.t mBP.While building interface between metal and mBP, minimum strain values obtained for Au(110), Ti(11-21) and Pd(110) are 1.4\%, 0.89\% and 1.8\% respectively. These cleaved surfaces are usually not found while building interface with 2D material, so we followed up with  the conventional cleaved surfaces for every metal.\cite{OPL:7959476} Following further the algorithm two different geometries are mapped and an intrinsic mean lattice mismatch value of 7.8\% with Au(111), 7.6\% with Ti(0001) and 2.4\% with Pd(111) is obtained. 

Fig 4 (a), (b) and (c) highlight the the most stable geometry for both the metal and semiconducting mBP using strain matching algorithm \cite{stokbro2013first} that is internal to ATK. The number of atoms in the whole configuration and the structure of mBP are different for different geometries of metal-mBP interface. This deviation is in accordance with the strain matching algorithm which aims to minimize the lattice mismatch considering all among the possible Bravias lattice configurations. The strain depicted is the minimum value of mean strain corresponding to each combination of metal and mBP. Though the value of strain is pretty much high for Au and Ti, we stick to the most reported surfaces (111) and (0001) for both Au and Ti respectively.
\begin{figure}
\includegraphics[width =0.8\columnwidth] {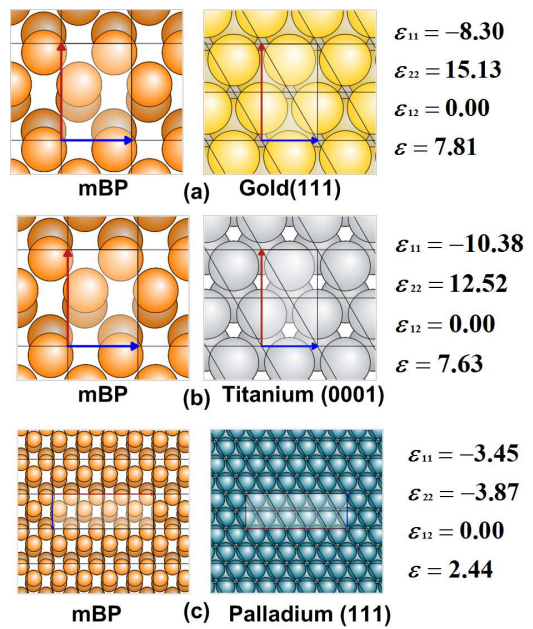}
\caption{\label{fig1}Most stable geometries obtained when two structures are interfaced in accordance with Strain Matching Algorithm. It can be seen here the number of atoms may increase or decrease as per the value of strain.The number is same for (a) Au (b) Ti and maximum for (c) Pd. The mean value of strain represented using $\epsilon$ is obtained by averaging all the three components of strain.    }
\end{figure}

\subsection{\label{sec:level2}DFT Simulations}
\subsubsection{\label{sec:level3}Optimization of Inter Layer Spacing}
Complying by the algorithm outlined in Fig 1, the next step is to examine the various analysis performed using DFT simulations. The optimized distances for all the metal-mBP interface are evaluated using the BE calculations. After building the interface geometry, we ran total energy (TE) simulations pertaining to all geometries varying inter layer distance from 1.5\AA{} - 3.5\AA{}. The TE comprises exchange correlation energy, kinetic energy, electrostatic energy and entropy. The BE is calculated as per the formula TE (metal + mBP) - TE (metal)- TE (mBP). The BE of Au-mBP, Ti-mBP and Pd-mBP is -0.74 eV, -1.94 eV and -5.19 eV respectively. The optimized distance corresponding to minimum BE and the BE variation curve for all the metal-mBP interface is shown in Fig 5 (a) - (d). 
\begin{figure}
\includegraphics[width =0.8\columnwidth] {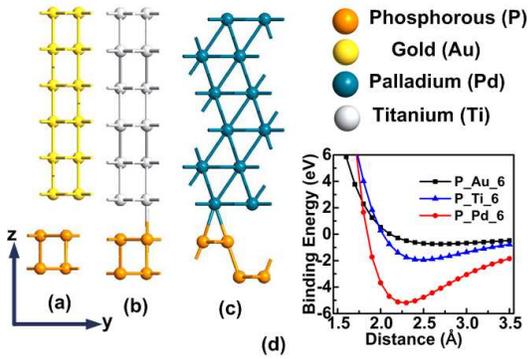}
\caption{\label{fig1}(a),(b),(c) Optimized geometry with inter layer distance corresponding to minima of BE. Bonding only exist for Ti and Pd, Au is stable without  bonding, (d) The BE curve for Au-mBP, Ti-mBP, Pd-mBP .}
\end{figure}
Here it is observed that the distance which is most energetically favorable is 2.7\AA{}, 2.5\AA{} and 2.3\AA{} for Au(111), Ti(0001) and Pd(111)-mBP interfaces respectively along the z-direction. Table I shows the covalent radius of each metal and its sum with the covalent radius of BP (1.07\AA{}) \cite{Cordero2008} including optimized distance obtained for each interface. Here it is observed that for Au-mBP interface the optimized distance is 2.7\AA{} which is 0.27\AA{} greater than sum of covalent radius of Au and BP.  In the configuration, Au atoms lie in the center of two phosphor atoms. This clearly indicates weak bonding between two atoms and minimum orbital overlap is expected here. Contrary to Au-mBP, the optimized distance for Ti-mBP and Pd-mBP is 2.5\AA{} and 2.3\AA{} which is less than the sum individual covalent radius  which is 2.67\AA{} and 2.46\AA{} respectively. The difference in sum of covalent radius and the optimized distance is less for Pd-mBP (0.16\AA{}) as compared to Ti-mBP (0.17\AA{}), so stronger bonding is expected in case of Pd and the same is reflected in Fig 5 (c). For Pd the both of the nearest Pd and P atoms bond whereas for Ti only one atom bond with P atoms. The Pd atom lies in the center of two P atoms while Ti atom lies directly above atom. As aforementioned, these geometries are in accordance with strain matching algorithm.
\begin{table}
\caption{\label{}Optimized Geometry for metal-mBP.}

\begin{ruledtabular}
\begin{tabular}{l*4{c}}
Metal & Covalent Radius\cite{Cordero2008} & Sum of metal+mBP & Optimized Distance\\
Au & 1.36 & 2.43 & 2.7\\
Ti & 1.60 & 2.67 & 2.5\\
Pd & 1.39 & 2.46 & 2.3\\
\end{tabular}
\end{ruledtabular}
\citetext{* The Covalent radius for BP is 1.07 \AA{}.\cite{Cordero2008} All the above has \AA{} units.}
\end{table}
\begin{figure}
\includegraphics[width =1.0\columnwidth] {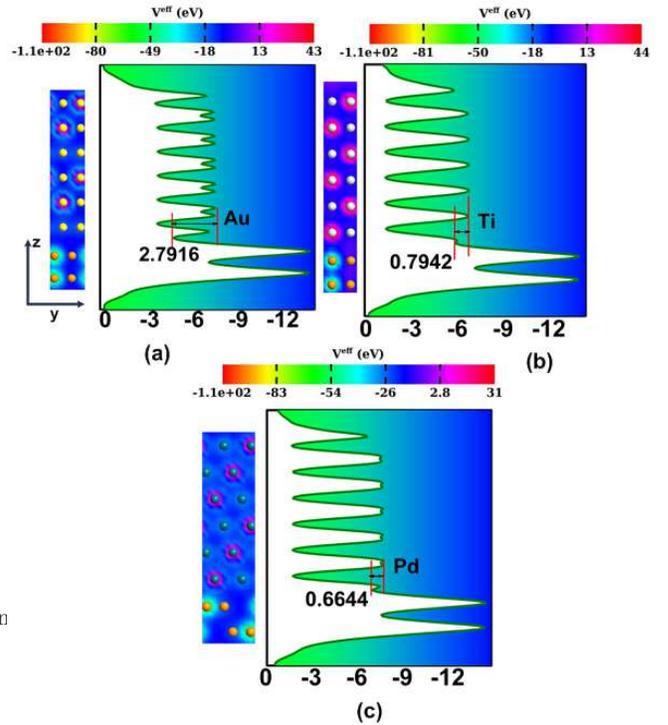}
\caption{\label{fig1}1D projection of effective potential curves for metal-mBP interface in z direction along with their 3D cut plane graphs on the left with atoms superimposed. Since the value is different for every metal, color bar of cut plane graphs are provided on the top for improving the readability. The value of barrier is amount of energy charge carrier must posses for the interface transport shown by red lines and are clearly stated for (a) Au-mBP (b) Ti-mBP and (c) Pd-mBP. x-axis represents the effective potential with units of eV. y-axis represents the distance along z-direction.}
\end{figure}

\subsubsection{\label{sec:level3}Interface Barrier Evaluation}
Henceforth, we proceed towards examining the electronic properties metal-mBP interface obtained using DFT simulations thus forming the second step of the algorithm (Fig 1). This step all together comprises eight different analysis, all of which leads to the final conclusion of interpretation of interface quality. For all the figures comprising the analysis,  we show mBP and  6 layers of metal atoms. 

The cut plane surfaces as well as the average effective potential(EP) plots along the z direction are depicted in Fig 6 (a) - (c). EP is described as the sum of 3 different potentials; V$^{eff}$[n] = V$^{H}$[n] + V$^{xc}$[n] + V$^{ext}$. The first two terms describe interactions with other electrons and can also be described by electron density. V$^{H}$[n] is the Hartree potential which is the result of mean field electrostatic interactions. Quantum mechanical nature of electrons gives rise to exchange-correlation potential V$^{xc}$[n]. The external potential contributed from ions potential and electrostatic interaction with external electrostatic field is represented by V$^{ext}$. \cite{Quantumwise} 

The tunnel barrier at the interface is calculated as difference between potential within the metal-mBP gap and the potential of the metal atom. The more higher the value, the more difficult it is for charge carrier to cross the barrier. In accordance with the results presented in Fig 6 (a) - (c) the  interface barrier is 2.7916 eV, 0.7942 eV and 0.6644 eV for Au-mBP, Ti-mBP and Pd-mBP respectively shown by red lines. Based on these values, it can be concluded that Au-mBP is a Schottky contact with a very less carrier injection efficiency. The values for Ti-mBP and Pd-mBP are pretty much low as compared to Au-mBP, resulting in a much better carrier injection efficiency than Au, still both of the above are not purely ohmic contacts as per the EP analysis on the interface and pose a small blockage for interface carrier travel.
 
\begin{figure}
\includegraphics[width =1.0\columnwidth] {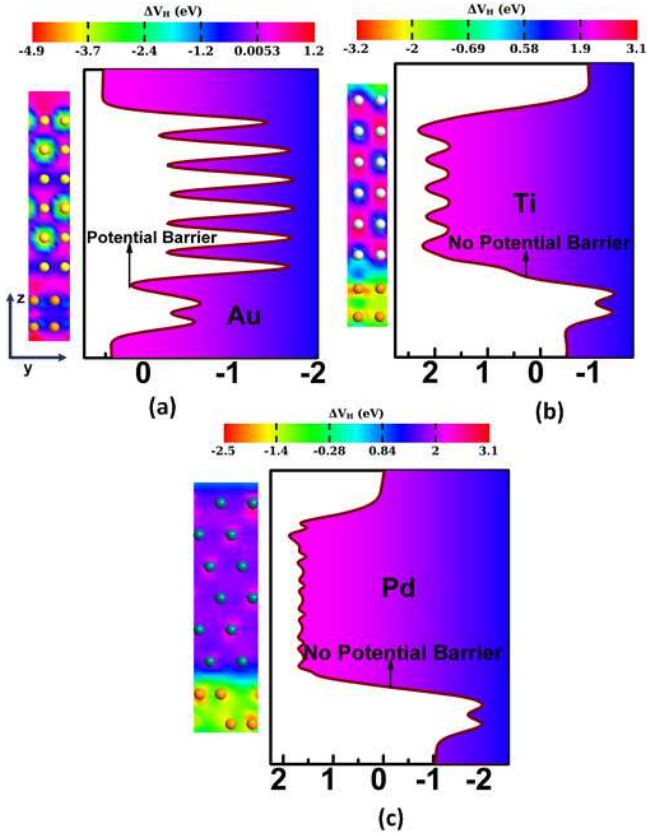}
\caption{\label{fig1} Electrostatic difference potential (EDP) of (a) Au-mBP, (b) Ti-mBP and (c) Pd-mBP   respectively, showing cut plane graphs with individual color bar for every metal and 1D projection along the z-axis. The potential is seen at the interface, where a black line shows an extra barrier at the metal mBP interface (a) and no barrier is indicated by a smooth line in (b) and (c) thus connecting the metal potential to mBP potential. x-axis represents the electrostatic difference potential with units of eV.y-axis represents the distance along z-direction.}
\end{figure}

Electrostatic difference potential (EDP) represents the difference between the electrostatic potential of self consistent valence charge density and electrostatic potential from a superposition of atomic valence densities.\cite{Quantumwise} It is also defined as the solution to the Poisson equation where charge density is the electron difference density which is explained in the section Interface Charge Transfer.  A negligible or zero value of EDP evinces no barrier at an interface. It is represented in Fig 7 (a) - (c) for Au, Ti and Pd metal contacts with mBP. It shows 3D cut plane geometries along with the 1D average plots integrated over x-y. From Figure 7 (a), it is clear that there is high potential for charge carriers at Au-mBP interface  and almost no barrier for Ti and Pd. This potential appears as an extra spike at the interface of Au and mBP (shown by black arrow in Fig 7(a)). This leads to a conclusion that there is more covalent bonding for Ti and Pd as compared to Au. While closely examining the curves, a very small protrusion is observed  in the line joining the potentials of Ti and mBP as compared to Pd-mBP where  the line very smooth. This indicates that the Pd bonding with mBP is higher as compared to the Ti.
\begin{figure}
\includegraphics[width =1.0\columnwidth] {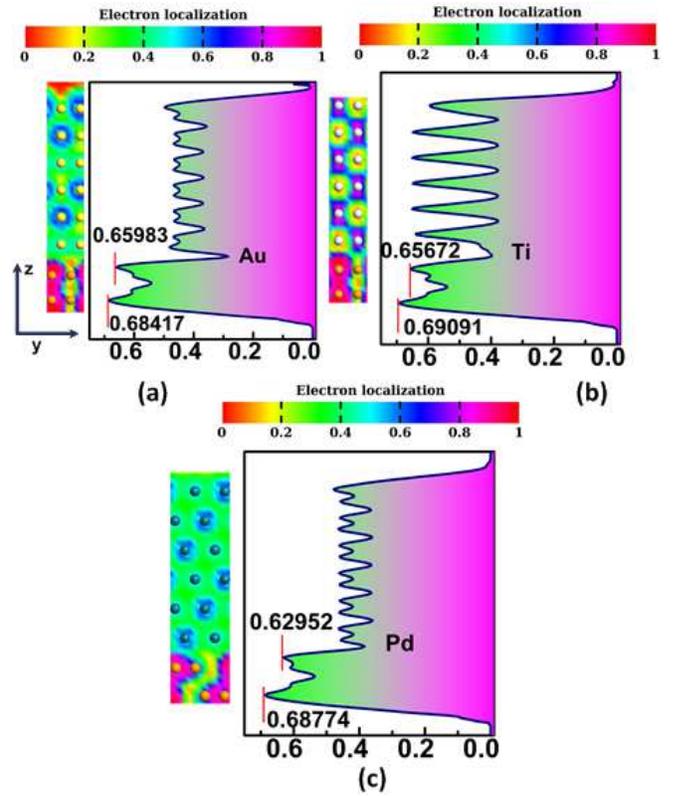}
\caption{\label{fig1}1D projection of electron localization function curves for metal-mBP interface along with their 3D cut plane graphs on the left with atoms superimposed. Here as well different color bars for  cut plane graphs are shown on the top of every combination of graphs. The value for lower P atom layer and upper P atom layer are scripted. A decreasing trend for the values of lower P atoms is observed for (a) Au-mBP (b)Ti-mBP and (c)Pd-mBP. x-axis represents the electron localization function.y-axis represents the distance along z-direction.}
\end{figure}

The averaged electron localization function (ELF) variation along z-direction with cut planes surfaces are depicted in Fig 8 (a) - (c). ELF as introduced by Becke and Edgecombe is a measurement of the probability of finding an electron in the adjacency of a reference electron. In the physical perspective, it quantifies the localization of the reference electron in space. Its value ranges from 0\textless ELF\textless 1, with the maximum limit ELF = 1 corresponding to prefect localization and ELF = 0.5 corresponding to electron-gas like pair probability.\cite{Becke1990,savin1992electron,silvi1994classification} Fig 8 (a) - (c) shows the ELF of P layer atoms near to and away from metal interface shown by red lines. The value is less for the layer closer to metal interface as compared to layer that is far away, for all the three metals.\cite{PhysRevB.90.125441} This shows that the closer P layer is less localized as compared to the outer P layer and which is quite apparent. If a comparison of the values of ELF for Au(0.65983), Ti(0.65672) and Pd(0.62952) are done, it is observed that the Au with the highest value of ELF has the maximum localization, which implies there is a minimum bonding between lower P atoms and lower Au atoms.  Between Ti and Pd, Pd has the lowest value indicating minimum localization and strong bonding between P atoms and Pd atoms. Above analysis can also be verified by the optimized geometry of metal-mBP interface as shown in Fig 5 (a) - (c).

Thus from the above three analysis, the fact is established that the carrier transfer is limited to a high extent by Au metal contact because of large tunnel barrier and minimum bonding at the interface with P atoms. These effects subsequently decrease for Ti and it is least for Pd.

\subsubsection{\label{sec:level3}Interface Charge Transfer}
The analysis for the describing interface charge transfer as drafted in Fig 1 are electron density (ED), electron difference density (EDD) and Mulliken population. They are the metrics which can effectively describe the charge measurements at specific locations. ED is the measure of quantum mechanical probability of locating electrons at particular positions in space.\cite{shusterman1997teaching} The higher the value of ED indicates more charge transfer has taken place at the interface. The averaged ED plots along z-direction with cut planes surfaces are shown in Fig 9 (a) - (c). The minimum values of ED at the interface are 0.11927 \AA{$^{-3}$}, 0.1295 \AA{$^{-3}$} and 0.21214 \AA{$^{-3}$} for Au-mBP, Ti-mBP and Pd-mBP respectively. Lowest value of ED are obtained for Au indicating weak bonding of Au and mBP. Whereas the highest value is obtained for Pd demonstrating strong covalent bonding and hence low resistance contact. The value for Ti is quite less in comparison to Pd but high in comparison to Au which shows an intermediate kind of bonding with respect to both Au and Pd.  
\begin{figure}
\includegraphics[width =1.0\columnwidth] {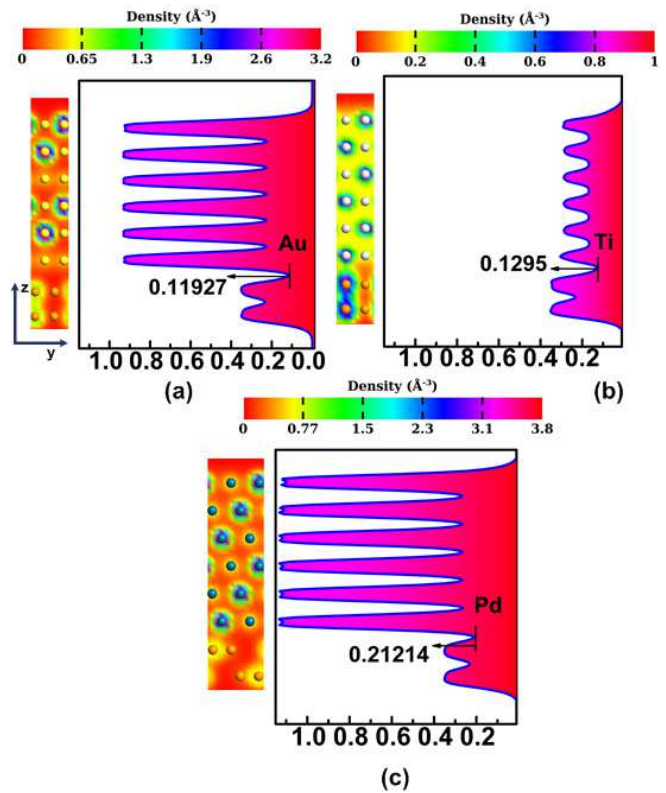}
\caption{\label{fig1}1D projection of electron density curves and their 3D cut plane graphs on the left for metal-mBP combination with proper atom placement. Various color bars are shown above each graph to avoid ambiguity in the presentation. Minimum electron density at the interface is displayed for (a) Au-mBP (b)Ti-mBP and (c)Pd-mBP. x-axis represents the electron density with units of \AA{}$^{-3}$.y-axis represents the distance along z-direction.} 
\end{figure}

Electron difference density (EDD) is the difference between the self consistent valence charge density and the superposition of atomic valence densities. A deviation between the total electron density of atoms and the density of neutral atoms is depicted by EDD.\cite{Quantumwise} It is that total charge density which can be further used to evaluate the EDP as explained earlier. When there is no interaction between the two different surfaces, means a negligible change in the electron density after the self consistent simulation at their boundary, hence it depicts a high EDD. On the other hand high interaction produces enough electron density, which when subtracted from the initial or neutral electron density can result in small values of EDD. The same is depicted by in Fig. 10 (a) - (c). The difference is shown by red arrows and is the highest for Au (0.0162\AA{$^{-3}$}) in comparison to both Ti (0.0117\AA{$^{-3}$}) and Pd (0.0042\AA{$^{-3}$}). The smallest is for Pd showing a maximum charge rearrangement between the mBP and Pd atoms. Hence a maximum charge interaction has taken place between Pd and mBP. This once again leads to the conclusion of maximum covalence between Pd and P atoms
\begin{figure}
\includegraphics[width =1.0\columnwidth] {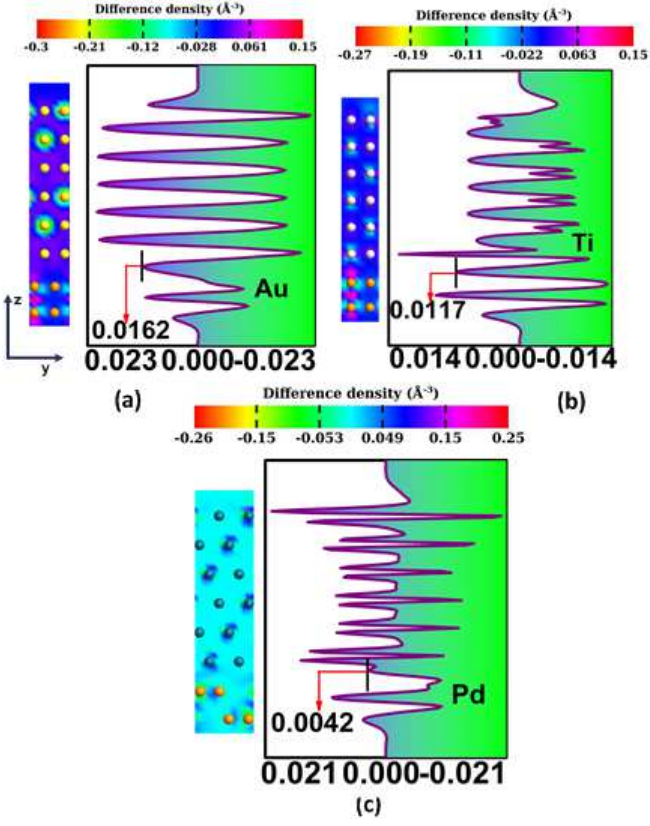}
\caption{\label{fig1} Red arrows show the EDD for (a)Au-mBP, (b)Ti-mBP and (c)Pd-mBP interface. The least value is for the Pd atoms describing maximum interaction. The 1D projection along z direction and cut plane graphs are presented side by side. x-axis represents the electron difference density with units of \AA{}$^{-3}$.y-axis represents the distance along z-direction.}
\end{figure}

\begin{figure}
\includegraphics[width =1.0\columnwidth] {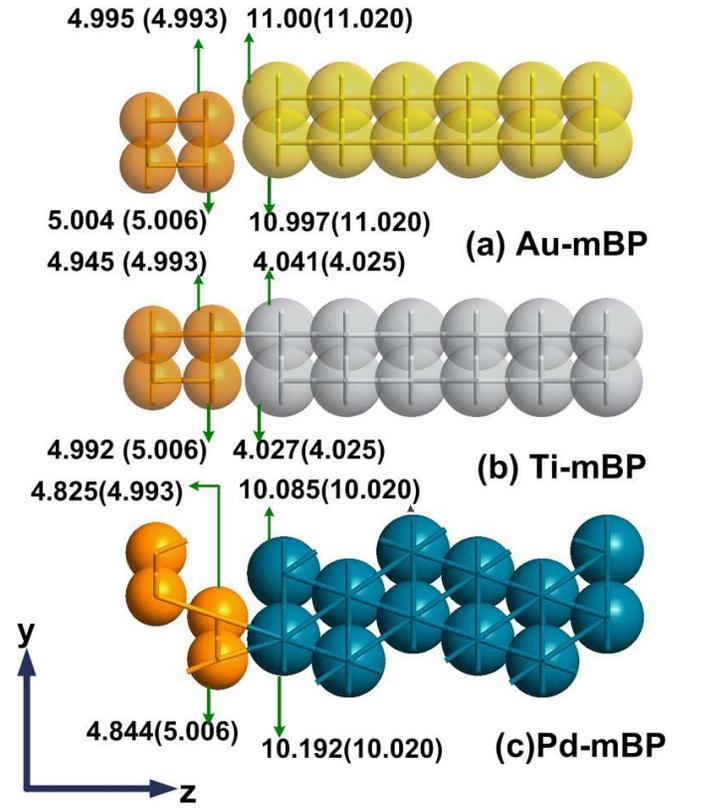}
\caption{\label{fig1}Mulliken Charge population of interfacial atoms for (a) Au-mBP,(b) Ti-mBP and (c) Pd-mBP are shown. The charge values inside and outside the brackets are before and after the the interface building of respective structures. The round spheres are the covalent radius of atoms and the lines indicate the bonds.  }
\end{figure}

Mulliken charges which appear due to Mulliken population analysis aids to provide net atomic populations and overlap populations of various atomic orbitals of different atoms.\cite{Mulliken1955} It is based on the Linear Combination of Atomic Orbitals (LCAO). High values of charges is indicative of large charge transfer between atoms. The Mulliken population of the atoms represented by Ball-Stick model after forming an interface and before interface, shown in brackets for each metal-mBP combination is shown in Fig 11 (a) - (c). The maximum \% change in the charge after interface formation is shown for both metal and P atom is presented in Table II. The change in the charge values for both P atom and Au atom is minimal for Au-mBP. However, this change increases for Titanium and is maximum for Palladium. The increased values of change in charges is indicative of the large amount of charge transfer has taken place between two atoms at the interface and thus confirming strong covalent bonding for Pd-mBP.

Based on the above three analysis pertaining to charge transfer at the interface, again it is confirmed that Pd has a lower resistance as comapred to both Ti and Au. 

\begin{table}
\caption{\label{}Maximum Change in Mulliken Charges of lower P atoms and lower metal atom.}

\begin{ruledtabular}
\begin{tabular}{l*3{c}}
Metal & \% Change in the P atom & \% Change in Metal atom\\
Au & 0.04 & 0.21 \\
Ti & 0.96 & 0.4 \\
Pd & 3.36 & 1.72 \\
\end{tabular}
\end{ruledtabular}
\end{table}

\subsubsection{\label{sec:level3}Interface Orbital Overlap}
\begin{figure}
\includegraphics[width =1.0\columnwidth] {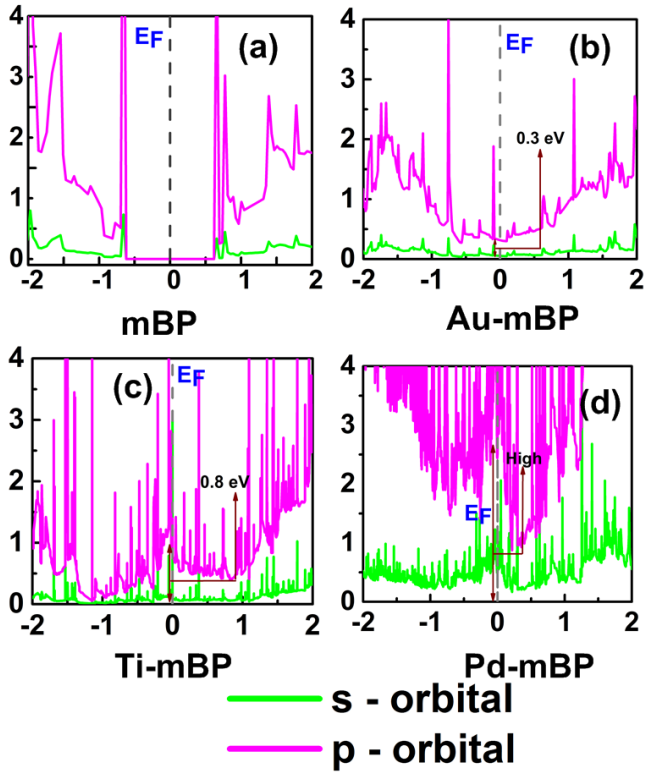}
\caption{\label{fig1}Projected density of states (PDOS)of individual P atoms showing the combined effect of s(green colour) and p (pink colour) orbitals in (a)mBP, (b)Au-mBP, (c)Ti-mBP and (d)Pd-mBP. E$_{F}$ is the Fermi Level shown by grey dashed lines. x-axis represent Energy in units of eV. y-axis represents the PDOS in arbitrary units. The shift of various orbitals near Fermi Level are shown by brown lines with arrows.}
\end{figure}
The last of the DFT simulations (Fig 1) include partial density of states (PDOS) and electronic band structure. The overlap between outermost atomic orbitals can be precisely studied using PDOS. A simple definition of density of states (DOS) depicts the number of states available that can be occupied by electrons at a particular energy level. Fig 12 (a), (b) - (d) show the cumulative PDOS of s and p orbitals of mBP and projected phosphorous atoms in the metal-mBP structure. The effect of p orbitals is more pronounced as compared to s-orbital in both the conduction and valence band regime of mBP and metal-mBP interfaces.  Due to a strong overlap of outermost orbitals at the interface the band gap nature of mBP is completely vanished for Ti and Pd and is preserved to some extent for Au-mBP. When compared with PDOS of pure mBP, it is observed that it is difficult to recognize exactly the VBM and CBM of mBP for the composite metal-mBP structures. The p-orbital shows an upward shift of 0.3 eV at the Fermi level as shown by brown arrows in Fig 12 (b). Closer to the Fermi Level we observe some empty states which do not appear for Ti-mBP and Pd-mBP. This clearly indicates that Au forms a resistive contact with mBP.  In case of Ti, the p-orbital shift at the Fermi level is 0.8 eV, which is very high and indicates metallization, thus showing the disappearance of semiconducting nature of the mBP. Peaks are observed closer to the Fermi level depicting less resistance for the carriers. Similar effects are seen for Pd-mBP as well where the peaks are higher and more dense as compared to Ti-mBP.The upward shift at Fermi level of p-orbital for Pd is also much higher as compared to Au and Ti. Hence, it can be deduced that Pd forms a puerly Ohmic contact for mBP.

\begin{figure}
\includegraphics[width =1.0\columnwidth] {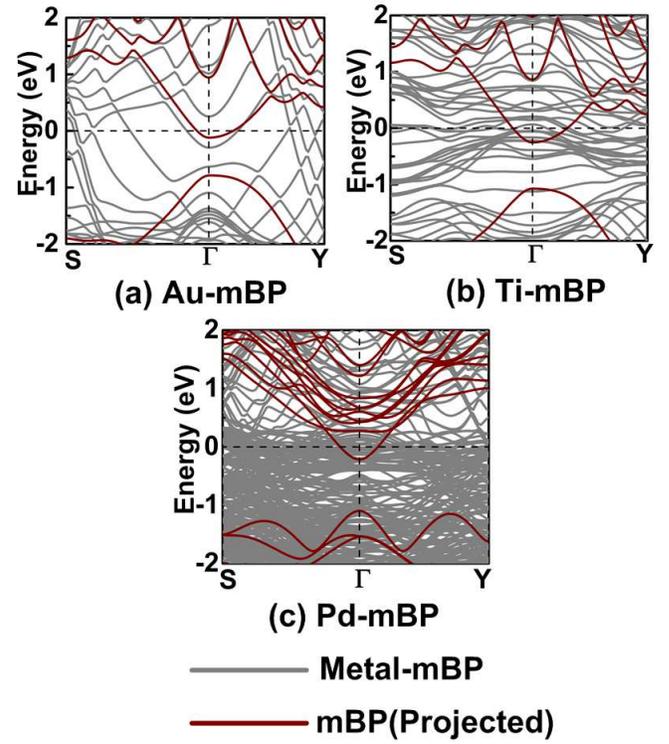}
\caption{\label{fig1}Electronic band structure showing comparison of metal-mBP bands with projected mBP bands for each contact metal (a) Au, (b) Pd and (c) Ti. Large mixing of bands suggest strong bonding. The shifting of mBP bands are visible in (a) for Au-mBP. The Fermi level is denoted by black dashed lines The symmetry points are different for every metal-mBP because of different dimensions of mBP unit cells.}   
\end{figure}

Finally we study the electronic band structure of the interface. Fig 13 (a) - (c) shows band structure of Au-mBP, Ti-mBP and Pd-mBP depicting  mixed metal-mBP band structure in grey colour with superimposed  projected band structure of mBP in brown  colour. As can be seen, the mBP band structure is perturbed by the presence of metal and it adapts a mixed metal-mBP nature. If the perturbation is quite high, the mBP conduction and valence bands are completely destroyed and a weak perturbation somewhat preserves the nature of mBP bands with a shift with respect to the Fermi Level.  Since the number of metal atoms are more (six layers are considered), the contribution by metal atoms is high as compared to the semiconductor in the electronic band structure and it can be examined for all the metals. From Fig 13 (a), for Au-mBP it is found that the conduction band (CB) and valence band (VB) of projected mBP at the gamma point are shifted downwards by crossing the Fermi level. If compared to the projected original electronic band structure of pure mBP (Fig.2(e)), it can be observed that due to interface orbital hybridization \cite{gong2014unusual} the bands have lost their original nature and have acquired a mixed metal-mBP character and do not entirely match the mBP CB and VB. For both Ti and Pd, the nature of mBP bands are completely destroyed at the gamma point as visible from Fig 13 (b)-(c). This depicts that both the metals have strong bonding with mBP and strong interaction between the two is observed. Hence presence of metal atoms alters the electronic structure of mBP and less interaction shifts the CB and VB with respect to the Fermi Level and strong interaction destroys the character of mBP bands describing formation of strong covalent bonds.

We also performed the metal-mBP contact analysis  using Grimme's DFT-D2 \cite{JCC:JCC20495} incorporating the effect of vander Waals interactions between the metal-mBP interface. It is found that the results remained unchanged and Au and Pd showed a Schottky and Ohmic nature respectively. 

Since unit cell of mBP is considered, the strain values obtained are high for Au(111) and Ti(0001).We performed a check for super cell of mBP till 10x10 repetitions. It is  observed that the minimum lattice mismatch is obtained for different super cell configurations of mBP with Au, Ti and Pd. The minimum lattice mismatch values obtained are for 1.7\% for  3x3, 4x4 and 5x5 supercell with Au(111); 1.6\% for 4x4 supercell with Ti (0001); and 1.5\% for 3x3 supercell with Pd(111). The repetitions are in x and y directions. Based on the above analysis it is found appropriate to interface only unit cell mBP with various metals. DFT simulations for mBP supercell interfaced with metals are also performed and no significant  variations in the final results are found.

All the above analysis are directly in correlation with the interlayer distance between the metal and mBP in the optimized geometry. For the present study and the reported systems, it is observed the more the distance, more likely are the chances for that particular metal to be a Schottky contact on the basis of interface properties as highlighted above and the less the distance, more interaction which leads to more bonding and the contact metal can likely be an Ohmic contact. So, we finally conclude that Au forms a purely Schottky contact and Pd forms an Ohmic contact with mBP confirmed by both DFT and DFT-D2. For Au,it is firmly believed that in presence of competent doping methods, it can also act as Ohmic contact.

\subsection{\label{sec:level2}Conclusion}
A systematic and organized study is conducted to assess the carrier injection efficiency various metal contacts (Au, Ti and Pd) to monolayer BP.  In optimized interface geometry, three quality assessment parameters: (i) the potential barrier, (ii) the electron transfer, and, (iii) the atomic orbital overlap are evaluated by employing the Kohn Sham density functional theory. These three parameters are in detail explained using effective potential, electrostatic difference potential, electron localization function, electron density, electron density difference, Mulliken population, density of states and electronic band structure to asses the Ohmic nature of the contacts. It is concluded that gold forms Schottky contact with a higher potential barrier at the interface and palladium offers purely Ohmic contact. The titanium contact demonstrates an intermediate behaviour.

\begin{acknowledgements}
{The work was supported by DST, Govt. of India, under grant 
no. SR/S3/EECE/0151/2012}
\end{acknowledgements}

\bibliography{Reference_Phospherene}

\end{document}